\documentclass[%
 reprint,
 superscriptaddress,
 nofootinbib,
 amsmath,amssymb,
 aps,
 prl,
]{revtex4-2}

\usepackage{graphicx}
\usepackage{dcolumn}
\usepackage{comment}
\usepackage{bm}
\usepackage{xcolor}
\usepackage[normalem]{ulem}

\newcommand{\V}{\mathcal{V}}

\begin{document}

\preprint{APS/123-QED}

\title{Stochastic Multiscale Reconstruction of Lagrangian Turbulence \\via Guided Diffusion Models}

\author{Conghui Wang}
\affiliation{Hangzhou International Innovation Institute, Beihang University, Hangzhou, Zhejiang 311115, PR China}

\author{Tianyi Li}
\email{tianyi.li@roma2.infn.it}
\affiliation{Department of Physics and INFN, University of Rome `Tor Vergata', Via della Ricerca Scientifica 1, 00133 Rome, Italy}

\author{Luca Biferale}
\affiliation{Department of Physics and INFN, University of Rome `Tor Vergata', Via della Ricerca Scientifica 1, 00133 Rome, Italy}

\author{Qinmin Zheng}
\email{zhengqinmin@buaa.edu.cn}
\affiliation{Hangzhou International Innovation Institute, Beihang University, Hangzhou, Zhejiang 311115, PR China}
\affiliation{Fluid Mechanics Key Laboratory of Ministry of Education, Institute of Fluid Mechanics, Beihang University, Beijing 100191, China}

\author{Michele Buzzicotti}
\affiliation{Department of Physics and INFN, University of Rome `Tor Vergata', Via della Ricerca Scientifica 1, 00133 Rome, Italy}

\author{Fabio Bonaccorso}
\affiliation{Department of Physics and INFN, University of Rome `Tor Vergata', Via della Ricerca Scientifica 1, 00133 Rome, Italy}

\date{\today}

\begin{abstract}
Lagrangian turbulence is characterized by intermittent, fat-tailed fluctuations and nontrivial correlations across temporal scales, making a quantitative description of its full multiscale probability distribution a longstanding challenge.
A particularly important question is whether unresolved fine-scale fluctuations can be inferred from coarse-grained trajectory information.
Here, we address this problem by sampling the conditional distribution of unresolved fluctuations using a diffusion-model prior conditioned on large-scale dynamics obtained through a wavelet-based coarse-graining of Lagrangian trajectories.
Using tracer trajectories from direct numerical simulations of homogeneous and isotropic turbulence at $Re_\lambda \simeq 310$, we show that the reconstructed signals recover scale-dependent intermittent statistics, including high-order structure functions, flatness, and local scaling exponents, together with cross-scale temporal correlations between resolved and unresolved fluctuations.
The method also reproduces the broad stochastic variability of intermittent acceleration fluctuations conditioned on the same coarse-grained trajectory, 
whereas Gaussian-process reconstructions in wavelet representation suppress rare events.
Our results show that small-scale Lagrangian intermittency can be modeled as a non-Gaussian conditional stochastic process constrained by coarse-scale dynamics and quantitatively reproduced through data-driven generative sampling.
\end{abstract}

\maketitle

\begin{figure}[!htbp] 
    \centering
    \includegraphics[
        width=0.5\textwidth,
        clip
    ]{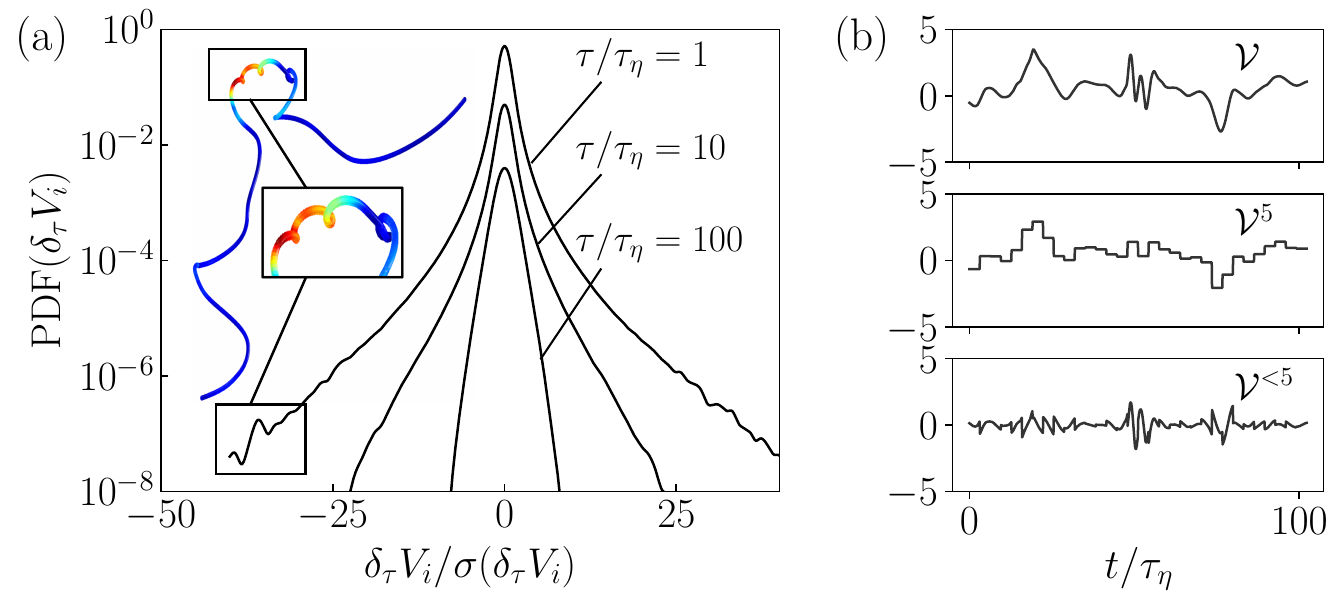}
    \caption{\label{fig:Multiscale characterization of original data} Multiscale characterization of Lagrangian turbulence. (a) Standardized PDFs of Lagrangian velocity increments at $\tau/\tau_\eta = 1,10,100$, with curves shifted vertically for clarity.
    Inset: representative vortex-trapping and extreme-acceleration event along a particle trajectory.
    (b) Original trajectory $\V$, coarse-grained large-scale component $\V^{5}$, and small-scale fluctuations $\V^{<5}$ obtained from a multiscale decomposition, shown for a single velocity component. All signals are normalized by the velocity standard deviation.}
\end{figure}

{\sc Introduction.}
Turbulence is a ubiquitous phenomenon in natural and engineering flows, characterized by strong nonlinearity, multiscale interactions, and spatio-temporal intermittency \cite{frisch1995turbulence,monin2013statistical,alexakis2018cascades}. 
While turbulent dynamics can be described in both Eulerian and Lagrangian frameworks, the latter---following individual fluid particles along their trajectories---provides a natural framework to access multiscale temporal correlations and intermittent extreme fluctuations \cite{la2001fluid,mordant2001measurement,mordant2002long,yeung2002lagrangian,toschi2009lagrangian}. 
Along Lagrangian trajectories, velocity increments and accelerations exhibit strongly non-Gaussian, fat-tailed statistics, a signature of intermittency across a wide range of temporal scales (Fig.~\ref{fig:Multiscale characterization of original data}a). 
A central challenge in turbulence is that these multiscale and intermittent fluctuations lack a quantitative statistical description. 
In particular, while coarse-grained representations of Lagrangian trajectories capture large-scale behavior, the non-Gaussian and multiscale statistical response of small-scale turbulent dynamics to large-scale motion remains largely unknown. 
This missing information plays a central role in intermittency, turbulence modeling, and subgrid-scale closures. 
This problem can be understood as the characterization and sampling of the conditional distribution of fine-scale fluctuations given coarse-scale observations. 
Despite their fundamental importance, such conditional statistics are currently not available, and no general framework exists to characterize and sample them consistently.

Over the past decades, Lagrangian turbulence has been studied through a variety of stochastic and phenomenological models, including Markovian and non-Markovian processes, as well as multifractal and multiplicative cascade approaches, which reproduce selected statistical features such as scaling laws and intermittency \cite{sawford1991reynolds,pope2011simple,viggiano2020modelling,zamansky2022acceleration,arneodo1998random,bacry2001multifractal,chevillard2019skewed,biferale1998mimicking,sinhuber2021multi}. 
However, these approaches do not provide a systematic framework to quantitatively characterize the full multiscale statistics of Lagrangian turbulence. As a result, generating realistic Lagrangian trajectories with correct multiscale behavior remains challenging. 
Although related stochastic interpolation approaches have been proposed \cite{friedrich2020stochastic,lubke2023stochastic}, it remains unclear whether fine-scale turbulent fluctuations can be inferred from coarse-grained observations in a statistically consistent manner that preserves multiscale correlations and intermittency.

In this Letter, we build on recent advances in generative modeling of Lagrangian turbulence \cite{li2024synthetic,guastoni2025new}, which demonstrated that data-driven diffusion models can generate trajectories with quantitatively accurate multiscale statistics. 
We introduce a framework to infer fine-scale fluctuations conditioned on coarse-grained (low-frequency) observations. 
By combining a learned prior with a conditional sampling strategy \cite{chung2022diffusion}, we reconstruct the fine-scale turbulent fluctuations in a manner consistent with both large-scale constraints and intermittent small-scale statistics.
\\
\\
{\sc Lagrangian turbulence data.}
We analyze Lagrangian velocity trajectories obtained from direct numerical simulations (DNS)
of homogeneous and isotropic turbulence at Taylor-scale Reynolds number 
$Re_\lambda \simeq 310$ \cite{biferale2023turb}. 
Each trajectory is sampled at temporal resolution $dt_s = 0.1 \tau_\eta$ 
over a duration $T \sim 0.7 \tau_L \sim 100 \tau_\eta$.
This setup spans a wide range of dynamically relevant time scales, 
from the dissipative scale $\tau_\eta$ to the integral scale $\tau_L$, 
covering more than two decades of temporal scales.
Velocity increments are defined as
\begin{equation}
\delta_\tau V_i(t) = V_i(t+\tau) - V_i(t),
\end{equation}
and exhibit strongly non-Gaussian statistics across these scales 
(Fig.~\ref{fig:Multiscale characterization of original data}),
with increasing intermittency toward smaller scales.
Further details on the simulations and data are provided in Appendix A.
\\
\\
{\sc Multiscale wavelet decomposition.}
We consider a Lagrangian velocity trajectory $\V(t)$ sampled at temporal resolution $dt_s$ and discretized into $K=2^N$ time points, which can be written as $\V=\{V_i(t_k)\}$ with $i=x,y,z$ and $k=1,\dots,K$. 
We introduce dyadic temporal scales labeled by $j=0,\dots,N$, with characteristic scale $2^j dt_s$, where larger $j$ corresponds to coarser resolution. 
We denote by $\V^j(t)$ the corresponding coarse-grained {\it approximation} of the signal, with $\V^0(t)$ the original trajectory. For simplicity, the following decomposition is written for a generic velocity component. Upon coarse-graining from level $j-1$ to $j$, the signal decomposes as
\begin{equation}
    \V^{j-1}(t) = \V^{j}(t) + D_j(t),
\end{equation}
where $D_j(t)$ denotes the scale-localized {\it detail} at temporal scale $2^j dt_s$. 
Iterating this decomposition yields
\begin{equation}
    \V(t) = \V^{J}(t) + \sum_{j=1}^{J} D_j(t) 
    \equiv \V^{J}(t) + \V^{<J}(t),
    \label{eq:wavelet_decomposition}
\end{equation}
where $\V^{<J}$ collects fluctuations at scales smaller than $2^J dt_s$. 
This construction corresponds to an orthogonal wavelet decomposition \cite{mallat2002theory}. 
Equivalently, the decomposition can be expressed in terms of approximation and detail coefficients $a_j(k)$ and $d_j(k)$ associated with each scale $j$, as detailed in Appendix B.
\\
\\
{\sc Stochastic multiscale reconstruction.}
The goal is to characterize and sample the fine-scale fluctuations conditioned on a coarse-grained observation:
\begin{equation}
P(\V^{<J} \mid \V^J ).
\label{eq:conditional_distribution}
\end{equation}
To illustrate the complexity of this task, Fig.~\ref{fig:Multiscale characterization of original data}(b) shows an example of a trajectory and its decomposition into approximation and detail components, highlighting the nontrivial cross-scale correlations underlying intermittency.
We employ an unconditional diffusion model trained on the present dataset
\cite{li2024synthetic,li2026deterministic},
which accurately reproduces the statistical properties of Lagrangian trajectories 
and defines an implicit prior distribution $P(\V)$ over trajectories. 
Sampling from this prior is performed through an iterative reverse diffusion process from Gaussian noise $\V_M$ to a trajectory sample $\V_0$, producing a sequence of states $\V_M,\dots,\V_m,\V_{m-1},\dots,\V_0$.
To approximate $P(\V^{<J}\mid \V^J)$, 
we guide the sampling process by enforcing consistency with the coarse-grained observation, following a diffusion posterior sampling (DPS) approach \cite{chung2022diffusion}.
Specifically, at each reverse diffusion step, the updated state $\V_{m-1}$ is corrected according to
\begin{equation}
\V_{m-1}
\leftarrow
\V_{m-1}
-
\xi
\nabla_{\V_m}
\|
\V^J
-
\widehat{\V}_{0|m}^J
\|^2,
\label{eq:dps_update}
\end{equation}
where $\widehat{\V}_{0|m}$ denotes the model estimate
of the clean trajectory given the current state $\V_m$,
and $\widehat{\V}_{0|m}^J$
its coarse-grained component, while $\xi$ controls the guidance strength.
Additional details on the proposed method,
including the selection of $\xi$,
are reported in Appendix C and Supplemental Material (SM).
This procedure can be interpreted as
an approximate Bayesian posterior sampling scheme
\cite{chung2022diffusion},
where the learned diffusion prior is conditioned
on the prescribed coarse-grained observation.
\\
\\
{\sc Results.}
We assess whether the proposed framework can reconstruct fine-scale statistics of intermittent Lagrangian turbulence from coarse-grained observations, while preserving their cross-scale correlations with the large-scale components. 
To this end, we reconstruct trajectories using both the proposed DPS method and a Gaussian process regression (GPR) baseline \cite{williams2006gaussian}, applied in the wavelet representation of the velocity trajectories (see Appendix D for details), and compare both against the DNS ground truth. 
The guidance strength $\xi$ in Eq.~(\ref{eq:dps_update}) is selected using a validation dataset by jointly considering pointwise reconstruction error and higher-order statistical properties.
We consider two coarse-graining levels, $J=5$ and $J=7$, corresponding to truncation scales of a few Kolmogorov times $\tau_\eta$ (approximately $3\tau_\eta$ and $10\tau_\eta$, respectively). 
In the main text we present results for $J=5$, where the truncation lies close to the transition between inertial and dissipative scales, while results for $J=7$, corresponding to a truncation within the inertial range, are reported in the SM. 
To quantify the statistical fidelity of the reconstruction, we adopt three complementary diagnostics, probing single-scale statistics, cross-scale correlations, and stochastic variability of intermittent fluctuations.

\begin{figure}[!h] 
    \centering
    \includegraphics[
        width=0.5\textwidth,
    ]{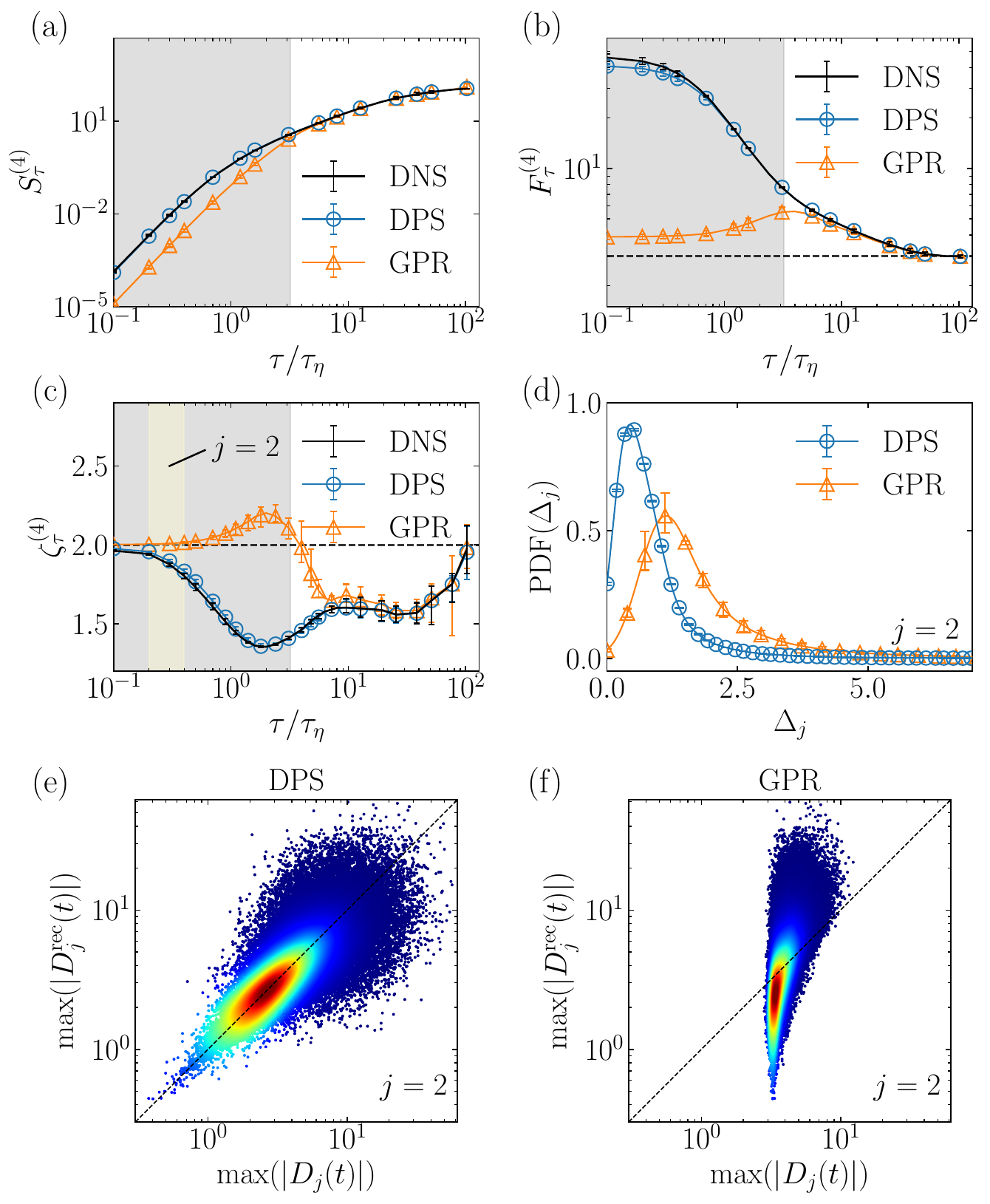}
    \caption{\label{fig:reconstructed Statistical Characteristics of level5} Comparison of single-scale statistics between DNS data and trajectories reconstructed using DPS and GPR for the $J=5$ truncation configuration, where the grey region indicates the scales removed by coarse-graining and targeted for reconstruction.
    (a) Fourth-order structure function $S_\tau^{(4)}$; (b) Fourth-order flatness $F_{\tau}^{(4)}$; (c) Local scaling exponent $\zeta_\tau^{(4)}$. Error bars indicate the spread among the three velocity components. The highlighted region in (c) corresponds to the detail scale $j=2$. (d) PDF over trajectories of the normalized reconstruction error $\Delta_j$ evaluated at $j=2$. 
    (e,f) Scatter plots of the maximum amplitude of the detail component $D_j$ at $j=2$, comparing DNS and reconstructed trajectories obtained using (e) DPS and (f) GPR. Axes are normalized by the standard deviation of $D_j$ from DNS. Warmer colors indicate higher probability density.
    }
\end{figure}

\noindent {\it 1. Single-scale statistics.}
We first perform a scale-by-scale validation based on the structure functions
\begin{equation}
S_\tau^{(p)} = \langle (\delta_\tau V_i)^p \rangle,
\end{equation}
where $\langle \cdot \rangle$ denotes averaging over time and trajectories, and the component index $i$ is omitted assuming isotropy.
We further consider the fourth-order flatness and the corresponding local scaling exponent, defined as
\begin{equation}
    F_{\tau}^{(4)} = \frac{S_{\tau}^{(4)}}{(S_{\tau}^{(2)})^{2}}, \qquad
    \zeta^{(4)}_\tau = \frac{d \log S_{\tau}^{(4)}}{d \log S_{\tau}^{(2)}}.
\end{equation}
These quantities provide stringent tests of multiscale intermittency \cite{arneodo2008universal}.
Fig.~\ref{fig:reconstructed Statistical Characteristics of level5} compares the reconstructed statistics obtained with DPS and GPR against the DNS reference. 
The grey region indicates the range of small-scale fluctuations removed by the coarse-graining and targeted for reconstruction. 
In panel (a), GPR, which reproduces second-order statistics by construction (not shown), deviates from DNS in the fourth-order structure function, whereas DPS remains consistent with DNS across all time lags.
A similar behaviour is observed in panel (b) for flatness and in panel (c) for the local scaling exponent:
GPR underestimates the growth of flatness toward small scales and fails to capture the scale-dependent intermittency, while DPS reproduces both quantities in close agreement with DNS. 
In particular, GPR fails to capture the pronounced increase of flatness in the dissipative range, a key signature of intermittency that poses a stringent test for stochastic and phenomenological models.
In panel (d) we show the PDF over trajectories of the normalized $L_2$ error between the ground truth and the multiscale reconstruction, evaluated on the detail components at a given scale $j$, defined as
\begin{equation}
\Delta_j = 
\frac{K^{-1}\sum_t (D_j(t) - D_j^{\mathrm{rec}}(t))^2}
{\langle D_j^2 \rangle},
\end{equation}
where $D_j$ and $D_j^{\mathrm{rec}}$ denote the ground truth and reconstructed detail components, respectively, and $\langle D_j^2 \rangle$ denotes the mean energy at scale $j$, averaged over time and trajectories.
Results are shown for $j=2$, corresponding to small-scale fluctuations at $\tau \sim 0.2$--$0.4\,\tau_\eta$. 
The corresponding scale range is highlighted in panel (c).
The error distribution for DPS is concentrated at small values and shifted toward lower errors compared to GPR.
To further assess reconstruction accuracy at the same scale, we consider the statistics of extreme fluctuations. 
Panels (e,f) compare, for each trajectory, the maximum amplitude of the detail component $D_j$ at $j=2$.
DPS exhibits a strong correlation with the ground truth across a broad range of fluctuations, whereas GPR collapses toward a narrow band and fails to reproduce large-amplitude values.
This shows that, beyond matching global statistics, DPS also achieves accurate reconstruction of fine-scale structures at the level of individual realizations, a nontrivial requirement for multiscale turbulent reconstruction.

\begin{figure}[!h] 
    \centering
    \includegraphics[
        width=0.35\textwidth,
        clip
    ]{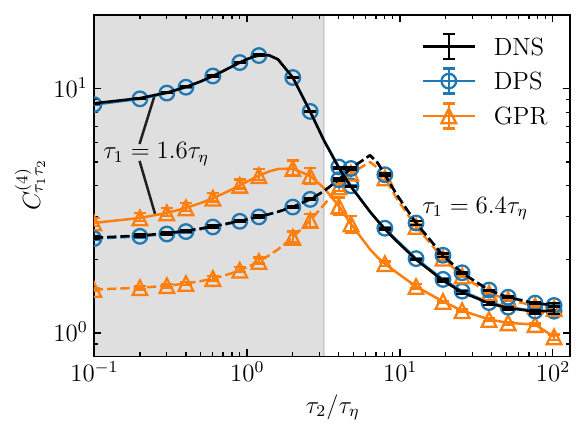}
    \caption{\label{fig:correlation} Fourth-order cross-scale correlation coefficient $C_{\tau_1\tau_2}^{(4)}$ for two representative values of $\tau_1$.
    Solid lines correspond to $\tau_1 = 1.6\tau_\eta$, within the reconstructed range, while dashed lines correspond to $\tau_1 = 6.4\tau_\eta$, within the resolved range. The grey region indicates the reconstructed scales.}
\end{figure}

\noindent {\it 2. Cross-scale correlations.}
We next assess whether the reconstruction captures cross-scale correlations between velocity fluctuations at different scales \cite{l1996fusion,benzi1998multiscale}.
To this end, we consider a two-time-lag correlation function based on velocity increments at $\tau_1$ and $\tau_2$:
\begin{equation}
C_{\tau_1 \tau_2}^{(4)} =
\frac{\langle (\delta_{\tau_1} V_i)^2 (\delta_{\tau_2} V_i)^2 \rangle}
{\langle (\delta_{\tau_1} V_i)^2 \rangle \, \langle (\delta_{\tau_2} V_i)^2 \rangle},
\end{equation}
where $i$ denotes a generic velocity component.
For $\tau_1 = \tau_2$, this quantity reduces to the fourth-order flatness defined above. 
We fix one time lag $\tau_1$ and scan the second lag $\tau_2$ across the resolved and reconstructed ranges.
Fig.~\ref{fig:correlation} shows $C_{\tau_1\tau_2}^{(4)}$ for two representative values of $\tau_1$, chosen in the reconstructed ($\tau_1 = 1.6\tau_\eta$) and resolved ($\tau_1 = 6.4\tau_\eta$) ranges. 
When both $\tau_1$ and $\tau_2$ lie in the resolved range, all methods recover the DNS correlations, as expected. 
Outside this regime, DPS remains in close agreement with DNS, whereas GPR exhibits significant deviations, especially as $\tau_2$ enters the reconstructed small scales (grey region).
This suggests that DPS captures cross-scale coupling between large and small scales.

\begin{figure}[!h] 
    \centering
    \includegraphics[
        width=0.5\textwidth,
        clip
    ]{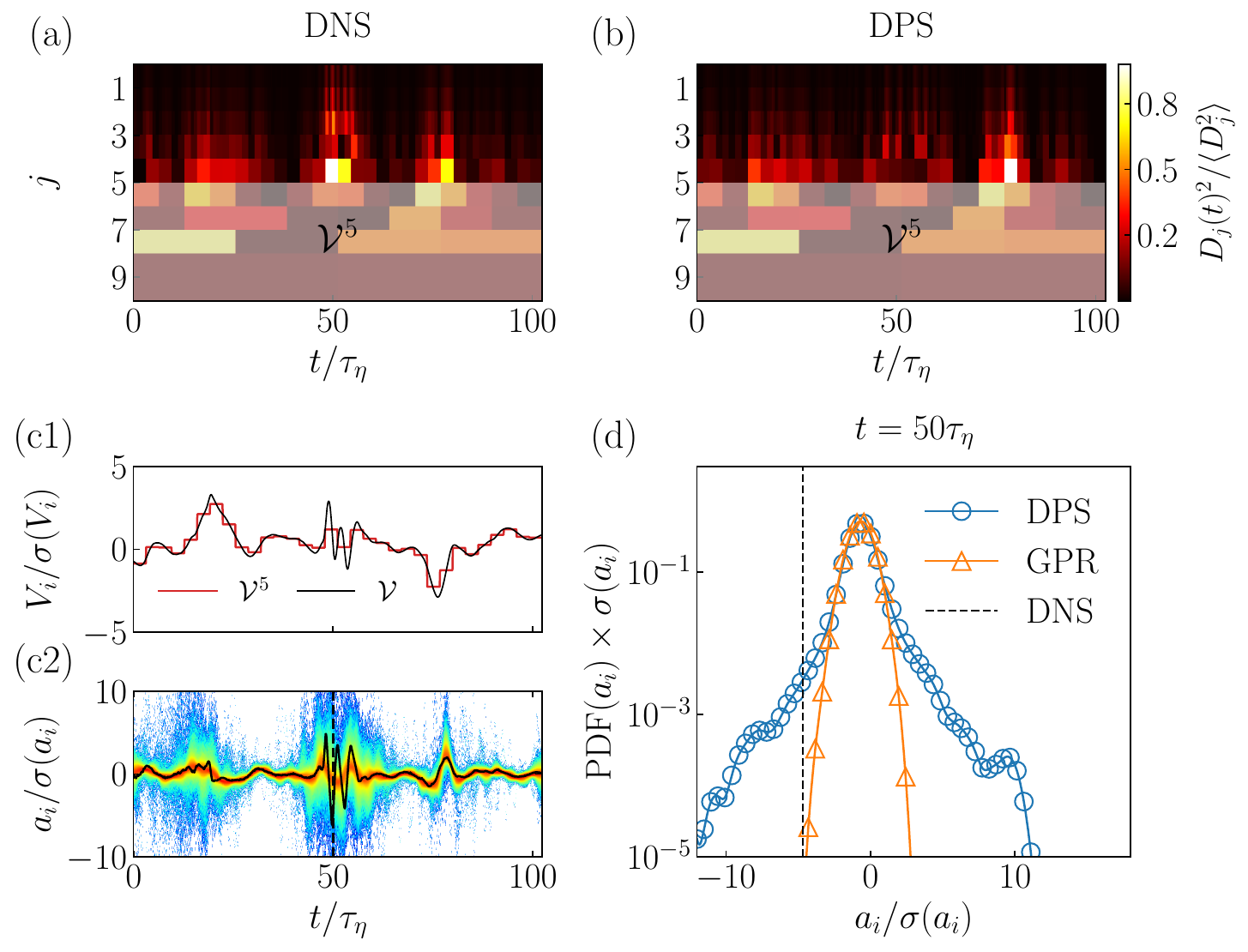}
    \caption{\label{fig:pdfs of a single trajectory}
Stochastic reconstruction of intermittent fluctuations conditioned on a fixed coarse-grained trajectory $\V^{5}$.
(a,b) Time--scale distribution of the normalized detail energy 
$D_j(t)^2/\langle D_j^2\rangle$
for a DNS trajectory and a corresponding DPS reconstruction.
The shaded region indicates the imposed coarse-grained component $\V^{5}$.
(c1) Imposed coarse-grained component $\V^{5}$ and corresponding original trajectory $\V$ for a single velocity component.
(c2) Conditional distribution of the acceleration component $a_i/\sigma(a_i)$ obtained from multiple DPS reconstructions conditioned on the trajectory shown in (c1).
The dashed line indicates the selected time $t=50\tau_\eta$ used in (d).
(d) Conditional PDFs of the acceleration component at 
$t=50\tau_\eta$
obtained from DPS and GPR reconstructions.
The dashed vertical line indicates the DNS value from the reference realization.
}
\end{figure}

\noindent {\it 3. Stochastic variability of intermittent fluctuations.}
We assess the stochastic variability of the reconstruction conditioned on a fixed coarse-grained trajectory $\V^J$. 
In contrast to the previous ensemble-averaged diagnostics, we fix a single realization of the large-scale component and analyze multiple stochastic reconstructions of the unresolved fluctuations.
The imposed coarse-grained trajectory for one velocity component is shown in Fig.~\ref{fig:pdfs of a single trajectory}(c1). 
Figs.~\ref{fig:pdfs of a single trajectory}(a,b) show the time--scale distribution of the normalized detail energy 
$D_j(t)^2/\langle D_j^2\rangle$
for a DNS trajectory and a corresponding DPS reconstruction.
The reconstructed scalogram exhibits realistic intermittent structures across scales.
Fig.~\ref{fig:pdfs of a single trajectory}(c2) shows the marginal PDF at each time of one acceleration component, a quantity dominated by the smallest temporal scales, obtained from multiple DPS reconstructions conditioned on the same coarse-grained trajectory.
The distribution is broad and strongly non-Gaussian, capturing a physically consistent range of small-scale fluctuations around the DNS value.
We further focus on a representative extreme fluctuation at 
$t=50\tau_\eta$,
indicated by the dashed vertical line in panel (c2).
Fig.~\ref{fig:pdfs of a single trajectory}(d) shows the corresponding conditional acceleration PDF at this time.
While DPS retains finite probability in the vicinity of the DNS event, GPR collapses to a much narrower distribution and strongly suppresses such rare fluctuations.
\\
\\
{\sc Conclusions.}
In summary, we have shown that unresolved small-scale Lagrangian turbulence can be quantitatively reconstructed as a conditional stochastic process constrained by coarse-scale dynamics. 
Conditioned on the coarse-grained component obtained via a wavelet-based decomposition, fine-scale fluctuations are sampled by guiding a learned generative prior.
The resulting reconstructions reproduce multiscale statistics, cross-scale correlations, and the broad non-Gaussian variability associated with intermittent fluctuations.
These results establish unresolved Lagrangian intermittency as a tractable conditional sampling problem and provide a quantitative framework for reconstructing its conditional statistics.
\\
\\
{\sc Acknowledgments.}
This research was supported by the National Natural Science Foundation of China under Grant No. U2441212 and 12225202, the Zhejiang Provincial Natural Science Foundation of China under Grant No. LZ25A020010, and the Research Funding of Hangzhou International Innovation Institute of Beihang University under Grant No. 2024KQ020. It was also supported by the European Research Council (ERC) under the European Union’s Horizon 2020 research and innovation programme Smart-TURB (Grant Agreement No. 882340) and by the Italian Ministry of University and Research (MUR) - FARE programme (No. R2045J8XAW).
L. B. and T. L. acknowledge INFN and CINECA for providing high-performance computing resources and support on the Leonardo supercomputer.
\\
\\
{\sc Data and Code Availability.}
The DNS Lagrangian trajectories used in this study are available from the open-access Smart-TURB portal at \url{http://smart-turb.roma2.infn.it}, in the TURB-Lagr repository \cite{biferale2023turb}. The code used to train and sample the unconditional diffusion-model prior is publicly available at \url{https://github.com/SmartTURB/diffusion-lagr}~\cite{diffusion_lagr_code}. The DPS sampling code developed for the conditional reconstructions in this work will be made publicly available upon publication.

\bibliography{apssamp}

\appendix

\section{Appendix A: Lagrangian dataset from Navier--Stokes simulations}

We use Lagrangian trajectories extracted from a direct numerical simulation of the three-dimensional incompressible Navier--Stokes equations in a periodic domain of size $L=2\pi$, forced at large scales to achieve a statistically stationary state \cite{biferale2023turb}.
Lagrangian particle positions $\bm{X}(t)$ and velocities $\bm{V}(t)$ are obtained by integrating passive tracers according to
\begin{equation}
\dot{\bm{X}}(t) = \bm{V}(t) = \bm{u}(\bm{X}(t), t),
\end{equation}
where $\bm{u}$ denotes the Eulerian velocity field obtained from the Navier--Stokes simulation.
A total of $N_p = 655{,}360$ trajectories are extracted from the simulation.
Each trajectory spans a duration
$T \sim 0.7\tau_L \sim 100\tau_\eta$
and is sampled at temporal resolution
$dt_s = 0.1\tau_\eta$,
resulting in $K=2^{10}=1024$ points per trajectory.
The dataset is divided into training,
validation, and test sets with a ratio of 8:1:1.
The validation set is used to select
the guidance strength $\xi$
in Eq.~(\ref{eq:dps_update}).

\section{Appendix B: Wavelet decomposition}\label{sec:wavelet}

We summarize the wavelet decomposition used in the main text (see Eq.~(\ref{eq:wavelet_decomposition})).
For simplicity, we write the decomposition
for a single velocity component;
the three-dimensional case is treated componentwise.
The coarse-grained approximation and the detail components are defined as
\begin{equation}
\begin{aligned}
\V^J(t) &= \sum_k a_J(k)\,\phi_{J,k}(t), \\
D_j(t) &= \sum_k d_j(k)\,\psi_{j,k}(t),
\end{aligned}
\label{eq:wavelet_coefficients}
\end{equation}
where $\phi_{J,k}$ and $\psi_{j,k}$ denote the scaling and wavelet basis functions, respectively \cite{mallat2002theory}, and $k$ labels the temporal location of the coefficients. 
At each scale $j$, the number of coefficients is $K_j = K / 2^j$.
In this work we employ the Haar wavelet basis; other choices would not affect the generality of the framework.

\section{Appendix C: Diffusion posterior sampling (DPS)}

We train an unconditional diffusion model
to define a learned prior distribution
$P_\theta(\V)$
over Lagrangian velocity trajectories
\cite{li2024synthetic,li2026deterministic}.
Trajectory generation is performed through
an iterative reverse stochastic process
over a total of $M$ diffusion steps,
\begin{equation}
\V_M
\rightarrow
\V_{M-1}
\rightarrow
\cdots
\rightarrow
\V_0,
\end{equation}
starting from Gaussian noise
$\V_M \sim \mathcal{N}(0,\mathbf{I})$,
where $\V_0$ corresponds to the generated trajectory
$\V$.
Each transition from step $m$ to $m-1$
is sampled from the learned transition probability
parameterized by neural network weights $\theta$:
\begin{equation}
\V_{m-1}
\sim
p_\theta(\V_{m-1} \mid \V_m)
\label{eq:reverse_transition}
\end{equation}
The diffusion architecture follows
the U-Net framework introduced in
Ref.~\cite{li2024synthetic}.
Additional implementation details are reported in the SM.
At each DPS sampling step,
the reverse transition in
Eq.~(\ref{eq:reverse_transition})
is followed by the correction step
introduced in Eq.~(\ref{eq:dps_update}).
The quantity
$\widehat{\V}_{0|m}$
corresponds to the model prediction
of the clean trajectory from
the intermediate state $\V_m$,
and can be interpreted as an approximation
of the posterior mean \cite{chung2022diffusion}
\begin{equation}
\widehat{\V}_{0|m}
\approx
\mathbb{E}_{p(\V_0|\V_m)}
\left[
\V_0
\mid
\V_m
\right].
\end{equation}
The correction term in
Eq.~(\ref{eq:dps_update})
therefore guides the reverse stochastic dynamics
toward trajectories whose coarse-grained component
matches the prescribed large-scale observation.

\section{Appendix D: Gaussian process regression (GPR)}

The GPR baseline models
the reconstruction problem
defined in Eq.~(\ref{eq:conditional_distribution})
in terms of the wavelet representation
introduced in Eq.~(\ref{eq:wavelet_coefficients}).
The wavelet decomposition is applied componentwise, and the resulting
coefficients of the three velocity components are considered jointly in the
GPR model.
We collect the coarse-grained coefficients into
\begin{equation}
\mathcal{A}_J =
\{a_{J,i}(k)\}_{i,k},
\end{equation}
and the detail coefficients into
\begin{equation}
\mathcal{D}_{<J} =
\{d_{j,i}(k)\}_{i,j,k},
\end{equation}
where $i=x,y,z$, $k=0,\dots,K_J-1$ for $\mathcal{A}_J$, and
$j=1,\dots,J$, $k=0,\dots,K_j-1$ for $\mathcal{D}_{<J}$.
We assume a joint multivariate Gaussian distribution
for the coarse-grained and detail coefficients,
\begin{equation}
\begin{bmatrix}
\mathcal{A}_J \\
\mathcal{D}_{<J}
\end{bmatrix}
\sim
\mathcal{N}
\left(
\begin{bmatrix}
\mu_A \\
\mu_D
\end{bmatrix},
\begin{bmatrix}
C_{AA} & C_{AD} \\
C_{DA} & C_{DD}
\end{bmatrix}
\right).
\end{equation}
Here $\mu_A=\langle \mathcal{A}_J \rangle$
and $\mu_D=\langle \mathcal{D}_{<J} \rangle$
denote the mean vectors estimated from the training dataset,
while
\begin{equation}
C_{AD}
=
\left\langle
(\mathcal{A}_J-\mu_A)
(\mathcal{D}_{<J}-\mu_D)
\right\rangle
\end{equation}
denotes the cross-covariance matrix between coarse-grained
and detail coefficients, with the remaining covariance
matrices defined analogously.
Given observed coarse-grained coefficients
$\mathcal{A}_J$,
the detail coefficients are reconstructed through
\begin{equation}
\begin{aligned}
P_{GPR}
(
\mathcal{D}_{<J}
\mid
\mathcal{A}_J
)
=
\mathcal{N}
(
&
\mu_D
+
C_{DA}C_{AA}^{-1}
(
\mathcal{A}_J-\mu_A
),
\\
&
C_{DD}
-
C_{DA}C_{AA}^{-1}C_{AD}
).
\end{aligned}
\end{equation}


\end{document}